\documentclass[prd,preprint,tightenlines,floatfix,
preprintnumbers,nofootinbib,eqsecnum,superscriptaddress]{revtex4}

\usepackage[T1]{fontenc}		
\usepackage{amsmath,amsfonts,amssymb,amstext,mathrsfs}
\usepackage{mathpazo}

\usepackage[dvips]{graphicx}
\usepackage{epsf,float}
\usepackage{revsymb}

\usepackage{soul}

\usepackage{dcolumn}
\usepackage{braket}
\usepackage{color,xcolor}
\usepackage{graphicx}
\usepackage{subfigure}
\usepackage{multirow}
\usepackage{tabularx}
\usepackage{pstricks}
\usepackage[section]{placeins}
\usepackage{booktabs}
\usepackage{array}

\usepackage{hyperref}

\newcommand{\ie}{{\it{i.e.}}}
\newcommand{\eg}{{\it{e.g}}}

\newcommand{\br}{\mbox{\boldmath $r$}}

\newcommand{\bb}{\mbox{\boldmath $b$}}

\begin{document}

\title{Dilepton Radiation in Heavy-Ion Collisions \\ at Small Transverse Momentum}


\author{Mariola K{\l}usek-Gawenda}
 \email{Mariola.Klusek@ifj.edu.pl}
\affiliation{Institute of Nuclear Physics Polish Academy of Sciences, Radzikowskiego 152, PL-31-342 Krak\'ow, Poland}

\author{Ralf Rapp}
 \email{rapp@comp.tamu.edu}
 \affiliation{Cyclotron Institute and Department of Physics and Astronomy, Texas A$\&$M University, College Station, TX 77843-3366, USA}

\author{Wolfgang Sch\"afer}
\email{Wolfgang.Schafer@ifj.edu.pl}
\affiliation{Institute of Nuclear Physics Polish Academy of Sciences, Radzikowskiego 152, PL-31-342 Krak\'ow, Poland}

\author{Antoni Szczurek
\footnote{Also at \textit{Faculty of Mathematics and Natural Sciences, University of Rzesz\'ow, Pigonia 1, PL-35-310 Rzesz\'ow, Poland}.}}
\email{Antoni.Szczurek@ifj.edu.pl}
\affiliation{Institute of Nuclear Physics Polish Academy of Sciences, Radzikowskiego 152, PL-31-342 Krak\'ow, Poland}

\today

\begin{abstract}
We study the invariant-mass distributions of dileptons produced in ultrarelativistic
heavy-ion collisions at very low pair transverse momenta,  $P_T\leq 0.15$\,GeV.  
Specifically, we investigate the interplay of thermal radiation with initial photon 
annihilation processes, $\gamma \gamma \to l^+ l^-$, triggered by the coherent 
electromagnetic fields of the incoming nuclei. For the thermal radiation, we employ
the emission from the QGP and hadronic phases with in-medium vector spectral functions 
which describes the inclusive excess radiation observed over a wide range of collision 
energies.  
For the coherent photon fusion processes, whose spectrum is much softer than for
thermal radiation, we employ initial fluxes from the Fourier transform of charge 
distributions of the colliding nuclei in the equivalent-photon approximation.  
We first verify that the combination of photon fusion, thermal radiation and final-state 
hadron decays gives a fair description of the low-$P_T$ dilepton mass spectra as recently 
measured by the STAR collaboration in $\sqrt{s_{NN}}$=200\,GeV Au+Au collisions for 
different centrality classes, including experimental acceptance cuts. The coherent 
contribution dominates in peripheral collisions, while thermal radiation shows a markedly 
stronger increase with centrality. We perform similar calculations at lower collision 
energies ($\sqrt{s_{NN}}$=17.3\,GeV) and compare to the acceptance-corrected dimuon excess 
spectra measured by the NA60 experiment at the CERN SPS; here, the contribution from 
photoproduction is subleading. We also provide predictions for the ALICE experiment at 
the LHC; the pertinent excitation function from SPS to LHC energies reveals a nontrivial 
interplay of photoproduction and thermal radiation.  
\end{abstract}


\maketitle


\section{Introduction}
\label{sec:intro}

Dilepton production in ultra-relativistic heavy-ion collisions (URHICs) has a long 
history as a probe of the hot QCD medium produced in these 
collisions~\cite{Feinberg:1976ua,Shuryak:1978ij}. 
To date, the observed excess radiation over final-state hadron decays has been firmly 
established as thermal radiation from the interacting fireball~\cite{Tserruya:2009zt,Specht:2010xu}. 
In the low-mass region, at invariant mass $M\lesssim 1\, \rm GeV$, hadronic radiation 
dominates revealing the melting of the $\rho$ resonance~\cite{Rapp:1997fs}, which 
indicates a transition to partonic degrees of freedom and is consistent with chiral 
symmetry restoration~\cite{Hohler:2013eba}. In the intermediate-mass region, 
quark-gluon plasma (QGP) radiation dominates~\cite{Rapp:1999zw,Ruppert:2007cr}, 
opening the possibility for direct measurements of its temperature~\cite{Rapp:2014hha}.  

On the other hand, recent measurement of dileptons in ultraperipheral heavy-ion
collisions (UPCs)~\cite{PHENIX,ALICE,ATLAS}, where the incoming nuclei do not touch and thus
no fireball is created, have also revealed a substantial amount of dilepton radiation
at both low and intermediate masses. This radiation is characterized by a very soft 
slope in pair transverse momentum, $P_T$, much steeper than for thermal radiation 
emitted from strongly interacting fireballs. A good description of the
ultraperipheral dilepton data can be achieved with photon fusion reactions 
using realistic fluxes generated by the electromagnetic (EM) field of the highly
relativistic incoming nuclei~\cite{KS2017}. 
The coherent EM fields are expected to also give a contribution 
in impact parameter configurations where the two ions collide.

The question then arises how the interplay of these two 
processes works out in peripheral heavy-ion collisions, where 
thermal radiation is much suppressed
compared to central collisions while the coherent photon emission is still appreciable.
Recently, two of us have shown that $J/\psi$ photoproduction in UPCs gives a significant
contribution to the low-$P_T$ production yield in semi-central collisions at the 
LHC~\cite{Klusek-Gawenda:2015hja}, in agreement with the ALICE results for different 
centralities~\cite{ALICE-jpsi}. Similar findings 
have been reported in Ref.~\cite{Shi}.
Recently, the STAR collaboration released new data for low-$P_T$ $e^+ e^-$ production 
in 200\,GeV Au+Au and U+U collisions over a large range of invariant mass and for different
centralities~\cite{Adam:2018tdm}. In that work, initial model comparisons were conducted
using contributions from thermal radiation~\cite{Rapp:2014hha} plus hadronic final-state 
decays plus photon fusion contributions from two different 
approaches~\cite{Klein:2018cjh,Zha:2018ywo}. The general trend was that photon fusion 
processes dominate, and can explain, the low-$P_T$ yield ($P_T$$<$0.15\,GeV) for peripheral 
collisions, although significant differences in the predicted yields are present.
On the other hand, thermal radiation plus the hadronic decay ``cocktail" increase much 
more rapidly with centrality and dominate the yield for $P_T$$>$0.2\,GeV at all centralities.    
A recent transport calculation~\cite{Song:2018dvf} confirmed that hadronic sources 
(thermal emission plus cocktail) cannot explain the low-$P_T$ excess observed by the 
STAR in peripheral Au-Au collisions. No photoproduction processes were considered in 
their study.

In the present paper we follow up on the above question by combining thermal radiation
with photon fusion processes for low-$P_T$ dilepton production. Quantitatively disentangling
the two contributions in this regime provides important tests for either one, especially 
if systematic centrality and collision energy dependences can be established.
After benchmarking our combined results against the STAR data at 200\,GeV, we therefore
expand our analysis to both lower and higher energies. For the former, we focus on In-In
collisions at SPS energy ($\sqrt{s_{NN}}$=17.3\,GeV) where we can test our predictions against 
the high-precision NA60 data~\cite{Arnaldi:2008fw,Specht:2010xu}; in fact, a low-$P_T$ excess 
was observed in these data which has not been fully explained to date.    

The paper is organized as follows. In Sec.~\ref{sec_sources} we briefly review the mechanisms 
for coherent photon fusion (Sec.~\ref{ssec_gamgam}) and thermal radiation (Sec.~\ref{ssec_thermal}).  
In Sec.~\ref{sec_spec} we apply them to low-$P_T$ dilepton invariant-mass spectra as measured
at RHIC (Sec.~\ref{ssec_rhic}) and the SPS (Sec.~\ref{ssec_sps}), make predictions for the LHC 
(Sec.~\ref{ssec_sps}), and compute an excitation function (Sec.~\ref{ssec_excit}). In
Sec~\ref{sec_concl} we conclude.   

\section{Dilepton Sources}
\label{sec_sources}
In this section we recall the main ingredients to the two main dilepton sources considered
in this work, \ie, the coherent initial photon fusion reactions in Sec.~\ref{ssec_gamgam} and
thermal radiation in Sec.~\ref{ssec_thermal}.

\subsection{Initial Photon-Photon Fusion Mechanism}
\label{ssec_gamgam}

The main ingredient for the photon-photon fusion mechanism is 
the Weizs\"acker-Williams flux of photons for an ion of charge $Z$ moving
along impact parameter $\bb$ ($b = |\bb|$) with the Lorentz-boost parameter $\gamma$. 
With the nuclear charge form factor $F_{\rm em}$ as an input the flux can be calculated 
as~\cite{Bertulani:1987tz, Baur:2001jj}
\begin{eqnarray}
N(\omega,b) 
&&= {Z^2 \alpha_{\rm EM} \over \pi^2} 
\Big| \int_0^\infty  dq_T {q_T^2   F_{\rm em}(q_T^2 + {\omega^2 \over \gamma^2} )   
	\over q_T^2 + {\omega^2 \over \gamma^2} } J_1(b q_T) \Big|^2\, ,  
\label{eq:WW-flux}
\end{eqnarray}
where $J_1$ is a Bessel function. We calculate the formfactor from the Fourier transform of the 
nuclear charge density, for which parameterizations are available in Ref.~\cite{DeJager:1987qc}.
 
%
 
The differential cross section for dilepton ($l^+ l^-$) production via $\gamma \gamma$ fusion at 
fixed impact parameter $\bb$ of a nucleus nucleus collision can then be written as
\begin{eqnarray}
{d \sigma_{ll} \over d\xi d^2\bb } =  
\int d^2\bb_1 d^2\bb_2 \, \delta^{(2)}(\bb - \bb_1 - \bb_2) N(\omega_1,b_1) N(\omega_2,b_2) 
{d \sigma(\gamma \gamma \to l^+ l^-; \hat s) \over d (-\hat t)} \ ,
\end{eqnarray}
where the phase space element is $d\xi = dy_+ dy_- dp_T^2$. Here, $y_\pm$, $p_T$ and $m_l$ are the single-lepton 
rapidities, transverse momentum and mass, respectively,
and
\begin{eqnarray}
\omega_1 = {\sqrt{p_T^2 + m_l^2} \over 2} \, ( e^{y_+} + e^{y_-} ) \  , \  
\omega_2 = {\sqrt{p_T^2 + m_l^2} \over 2} \, ( e^{-y_+} + e^{-y_-} ) \ , \ \hat{s} = 4 \omega_1 \omega_2 \ .
\end{eqnarray}
As can be seen from Eq.(\ref{eq:WW-flux}), the transverse momenta, $q_T$, of the photons have been 
integrated out, and dileptons are produced back-to-back in the transverse plane, \ie, the transverse
momentum $P_T$ of the pair is neglected. 

In UPCs the incoming nuclei do not touch, \ie, no strong interactions occur between them. 
In this case one usually imposes the constraint $b > 2 R_A$ when integrating over
impact parameter $b = |\bb|$.
Here we lift this restriction allowing the nuclei to collide. Then the final state will no longer contain the intact nuclei but the dileptons will be produced on top of the hadronic 
nuclear event characterized by an impact parameter $\bb$ (or range thereof).
Note that even for overlapping nuclei, $b < 2 R_A$, leptons are predominantly produced 
outside the overlap region, for $ b_{1,2} > R_A$. This situation is 
very different from the photoproduction heavy vector mesons~\cite{Klusek-Gawenda:2015hja} 
which tend to be produced ``inside" of one of the nuclei. 

Here we are interested in the dependence on centrality. The mass-differential dilepton yield 
from coherent photons in a centrality class ${\cal C}$ corresponding to an impact parameter 
range of $[b_{\rm min}, b_{\rm max}]$ is given by
\begin{eqnarray}
{dN_{ll}[{\cal C}] \over dM} = {1 \over f_{\cal C} \cdot \sigma^{\rm in}_{\rm AA}} 
\int_{b_{\rm min}}^{b_{\rm max}} db  \, 
\int d\xi \, 
\delta(M - 2 \sqrt{\omega_1 \omega_2}) \, 
{d \sigma_{ll} \over d\xi db }\Big|_{\rm cuts} \ , 
\end{eqnarray}
where we have indicated kinematic cuts on single-lepton variables as applied in experiment, and 
$f_{\cal C}$ is the fraction of inelastic hadronic events contained in the centrality class ${\cal C}$, 
\begin{equation}
f_{\cal C} = { 1 \over \sigma^{\rm in}_{\rm AA} } 
\int_{b_{\rm min}}^{b_{\rm max}} db \frac{d \sigma^{\rm in}_{\rm AA}}{d b}  \, .
\end{equation}
We determine $[b_{\rm min}, b_{\rm max}]$ and $\sigma^{\rm in}_{\rm AA}$ by using the optical 
Glauber model as
\begin{equation}
\frac{d \sigma^{\rm in}_{\rm AA}}{d b} = 2 \pi b (1 - e^{-\sigma^{\rm in}_{\rm NN} T_{\rm AA}(b)}) \; .
\end{equation}
The nuclear thickness function, $T_{\rm AA}(b)$, is obtained from
the convolution of nuclear density distributions for which we use standard Woods-Saxon 
profiles, $n_{\rm A}(r)$, 
\begin{equation}
T_{\rm AA}(b) = \int d^3 \vec{r}_1 d^3\vec{r}_2 \, 
\delta^{(2)}(\bb - \br_{1 \perp} - \br_{2 \perp}) \, n_{\rm A}(r_1) n_{\rm A}(r_2) \, .
\end{equation}

\subsection{Thermal Dileptons}	
\label{ssec_thermal}
Thermal dilepton radiation in URHICs is based on the idea that the abundant 
production of hadrons, together with strong re-interactions, leads to the formation 
of a locally equilibrated medium whose expansion can be described by relativistic
hydrodynamics. This idea is by now well established, on the one hand by the
success of hydrodynamic modelling in reproducing the transverse-momentum spectra 
of the produced hadrons~\cite{Shuryak:2008eq,Heinz:2013th,Gale:2013da}, and, on the 
other hand, by the observation and theoretical description of dilepton radiation 
that goes well beyond the final-state decays of the produced 
hadrons~\cite{Tserruya:2009zt,Specht:2010xu,Rapp:2013nxa,Rapp:2014hha}.  
The basic equation to compute dilepton invariant-mass spectra involves an integration
of the 8-differential emission rate over the space-time evolution of the expanding fireball, 
\begin{equation}
\frac{dN_{ll}}{dM} = \int d^4x \ \frac{Md^3P}{P_0} \ 
\frac{dN_{ll}}{d^4xd^4P} \ .
\end{equation}
where $(P_0,\vec P)$ and $M=\sqrt{P_0^2-P^2}$ are the 4-vector ($P=|\vec P|$) and invariant 
mass of the lepton pair, respectively.  The thermal emission rate can be expressed as
\begin{equation}
\frac{dN_{ll}}{d^4xd^4P} = -\frac{\alpha_{\rm EM}^2 L(M)}{\pi^3 M^2} \ 
f^B(P_0;T) \ {\rm Im}\Pi_{\rm EM}(M,P;\mu_B,T) \ ,
\label{rate}
\end{equation}
in terms of the Bose distribution function, $f^B$, and the EM spectral 
function, Im\,$\Pi_{\rm EM}$, depending on the local temperature, $T$, and baryon 
chemical potential, $\mu_B$, of the medium ($L(M)$ is a lepton phase-space factor which 
approaches one for $M\gg m_l$). The fireball medium generally goes through both
QGP and hadronic phases; for the respective spectral functions we employ in-medium 
quark-antiquark annihilation constrained by lattice-QCD~\cite{Rapp:2013nxa} and 
in-medium vector spectral functions in the hadronic sector~\cite{Urban:1999im}; the 
vector meson resonances strongly broaden in the medium and essentially melt at 
temperatures close to the pseudocritical one, providing a nearly smooth transition 
to the QGP rates. Different centrality classes for different colliding systems 
are characterized by the measured hadron multiplicities and
appropriate initial conditions for the fireball. Its expansion is modelled by a 
simple volume parameterization guided by hydrodynamic models, and the underlying 
equation-of-state is based on $\mu_B$=0 lattice-QCD results for the QGP smoothly
matched to a hadron resonance gas~\cite{Rapp:2014hha}. This approach is consistent
with available dilepton data from SIS~\cite{Galatyuk:2017ack} via 
SPS~\cite{Arnaldi:2008fw,Agakichiev:2005ai} to 
RHIC~\cite{Huck:2014mfa,Adamczyk:2015lme,Adare:2015ila} energies.

\section{Low-$P_T$ Dilepton Invariant-Mass Spectra}
\label{sec_spec}
We are now in position to combine the two sources described above and compare their
sum to low-$P_T$ measurements at RHIC (\ref{ssec_rhic}) and SPS (\ref{ssec_sps}) and make 
predictions for the LHC (\ref{ssec_lhc}) and their excitation function (\ref{ssec_excit}).

\subsection{Au-Au Collisions at RHIC}
\label{ssec_rhic}
\begin{figure}[!t]
        (a) \includegraphics[scale=0.375]{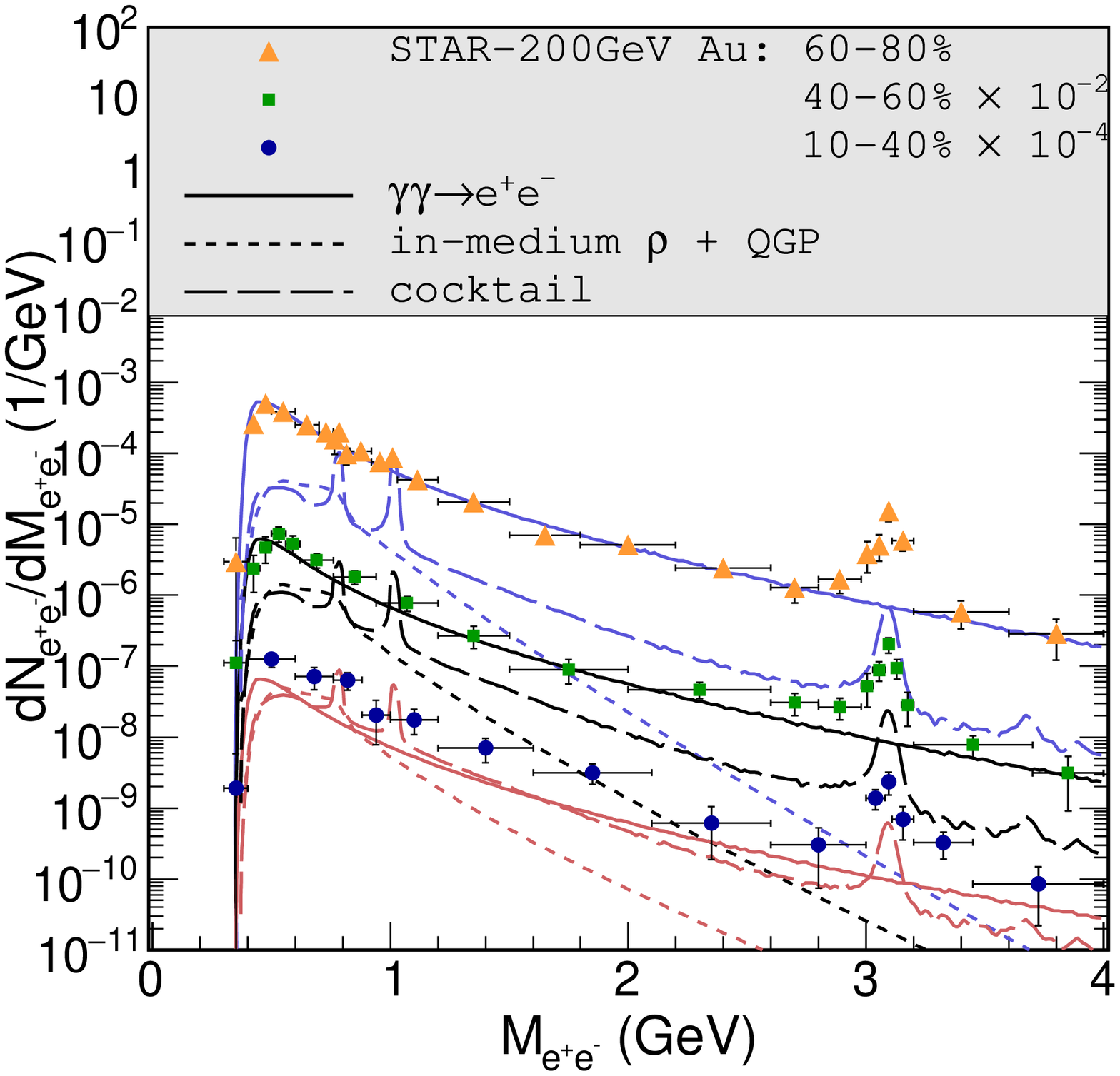}
        (b) \includegraphics[scale=0.375]{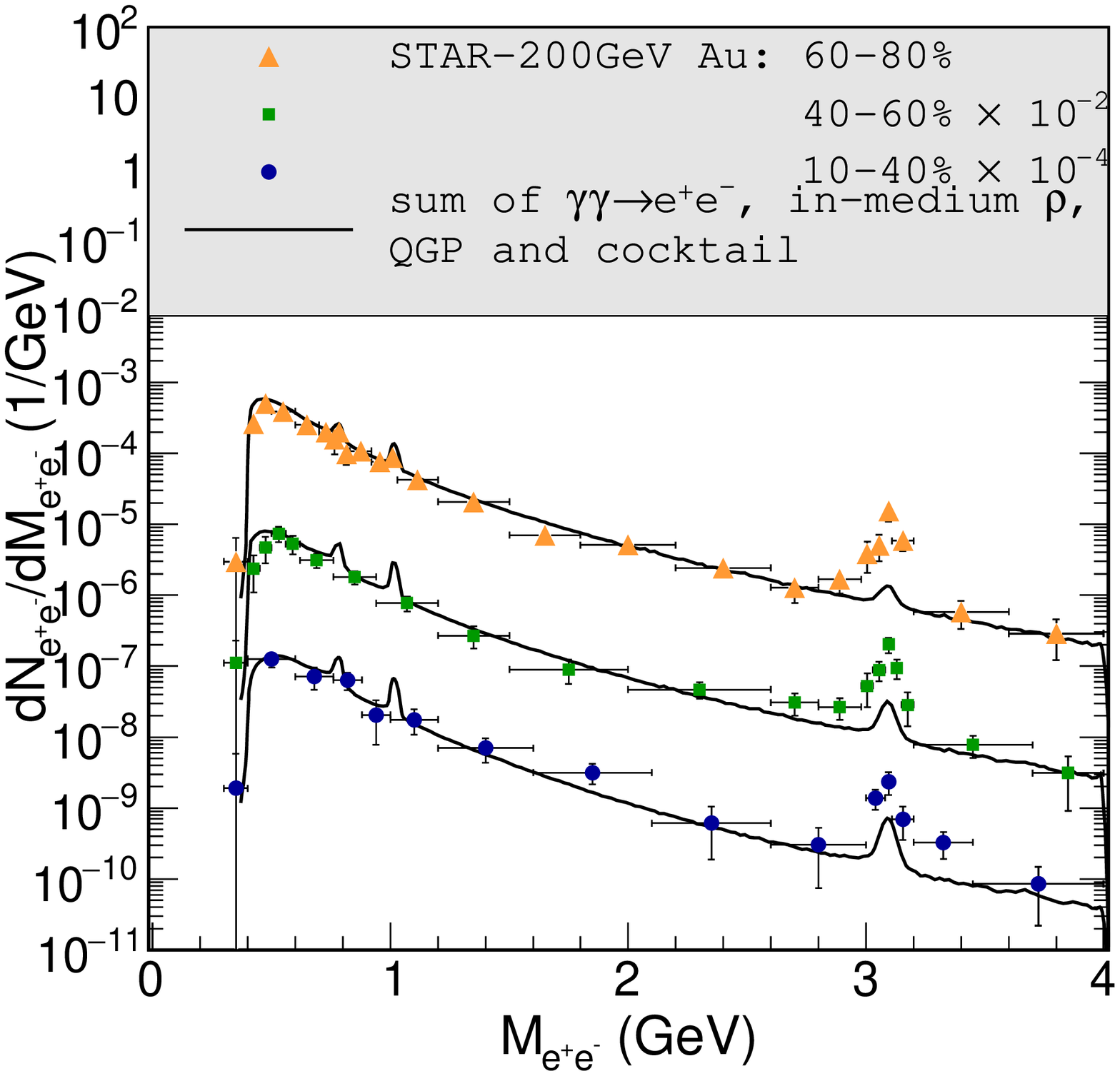}
        \caption{
(a) Dielectron invariant-mass spectra for pair $P_T$$<$0.15\,GeV in Au+Au ($\sqrt{s_{NN}}$=200\,GeV) 
collisions for three centrality classes calculated (including experimental acceptance
cuts, $p_{T}$$>$0.2\,GeV, $|\eta_e|$$<$1 and $|y_{e^+e^-}|$$<$1) for $\gamma \gamma$ fusion 
(solid lines), thermal radiation (dotted lines) and the hadronic cocktail (dashed lines).
(b) Comparison of the total sum (solid lines) to STAR data~\cite{Adam:2018tdm}.
}
\label{fig:photoproduction_vs_STAR}
\end{figure}
In Fig.~\ref{fig:photoproduction_vs_STAR} we show dielectron invariant-mass 
spectra for small pair $P_{T} <$ 0.15~GeV and three different 
centrality classes as selected in the STAR data: peripheral (60-80\%), semi-peripheral (40-60\%) and 
semi-central (10-40\%) collisions. We also include the experimental acceptance cuts on the
single-lepton tracks as applied by STAR, and take the cocktail contribution as provided by STAR~\cite{Adam:2018tdm} 
representing the final-state decays of the produced hadrons. 
In peripheral collisions the photon-photon contribution dominates while
in semi-central collisions all three contributions are of similar magnitude. 
Their sum yields a rather good agreement with the STAR data, except for the 
$J/\psi$ peak region. Our calculations only contain incoherent $J/\psi$ production, 
from binary nucleon-nucleon collisions;
we conjecture that the missing contribution is due to a coherent contribution
discussed, \eg, in Ref.~\cite{Klusek-Gawenda:2015hja}, which we do not further pursue here, as 
our focus is on the interplay with thermal radiation.

\subsection{In+In Collisions at the SPS}
\label{ssec_sps}
After benchmarking our approach with the STAR data, we now turn to the high-precision NA60 
data from In-In ($\sqrt{s_{NN}}$=17.3\,GeV) collisions at the SPS, for which no calculations of 
the coherent contribution are available to our knowledge. In addition, the dimuon $P_T$ 
spectra of the NA60 collaboration show a distinct hint for an enhancement at very low 
$P_T$ in various mass bins up to $M$$\simeq$1\,GeV that could not fully be explained by 
calculations of thermal radiation~\cite{Arnaldi:2008fw,Specht:2010xu}. 

\begin{figure}[!t]
	\includegraphics[scale=0.45]{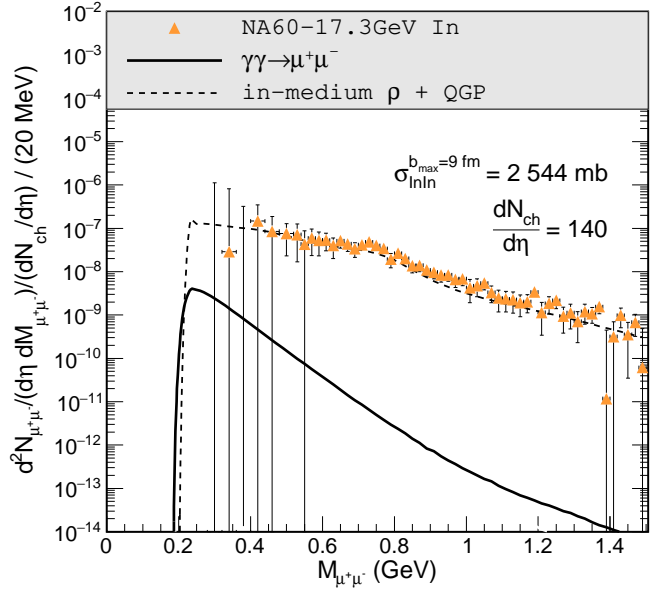}
\caption{Low-$P_T$ (<0.2\,GeV) acceptance-corrected dimuon invariant mass excess spectra in the rapidity range 3.3<$Y_{\mu^+\mu^-,LAB}$<4.2 for MB In+In ($\sqrt{s_{NN}}$=17.3 GeV) collisions at the SPS. Calculations for coherent
$\gamma \gamma$ fusion (solid line) and thermal radiation
(dashed line) are compared to NA60 data~\cite{Arnaldi:2008fw,Specht:2010xu}.
	}
\label{fig:NA60}
\end{figure}

In Fig.~\ref{fig:NA60} we show the results of our calculations for thermal radiation 
and the coherent $\gamma \gamma$ mechanism for minimum bias (MB) In-In collisions, in 
comparison to the acceptance corrected NA60 excess data (from which the cocktail has 
also been subtracted). Unlike for RHIC energies the $\gamma \gamma$ contribution is 
rather small and only plays some role at small dimuon invariant masses where the NA60 
data run out of precision. 

\subsection{Pb+Pb Collisions at the LHC}
\label{ssec_lhc}
\begin{figure}[!t]
	 \includegraphics[scale=0.28]{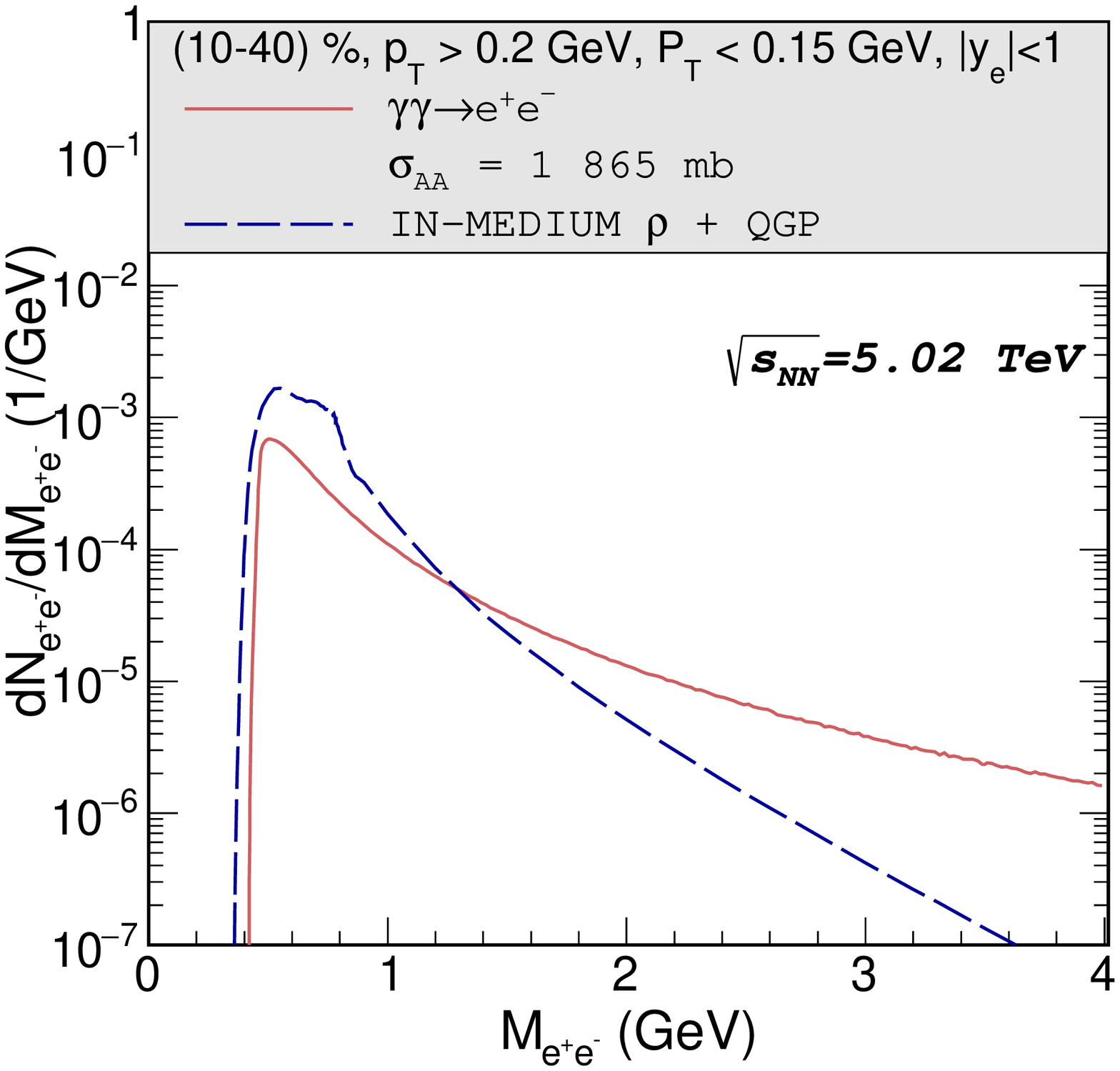}
\hspace{-0.4cm}
	 \includegraphics[scale=0.28]{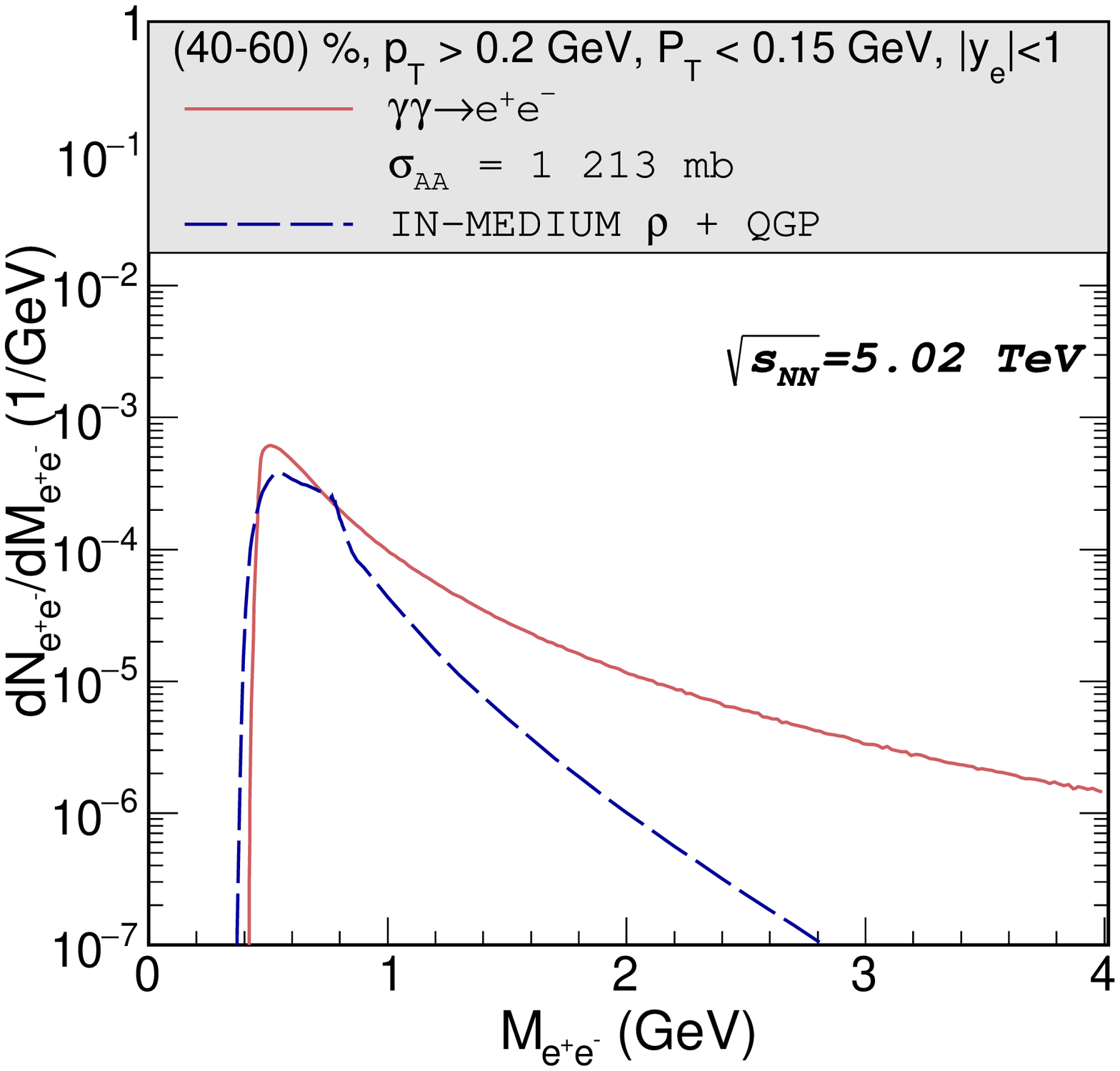}
\hspace{-0.4cm}
	 \includegraphics[scale=0.28]{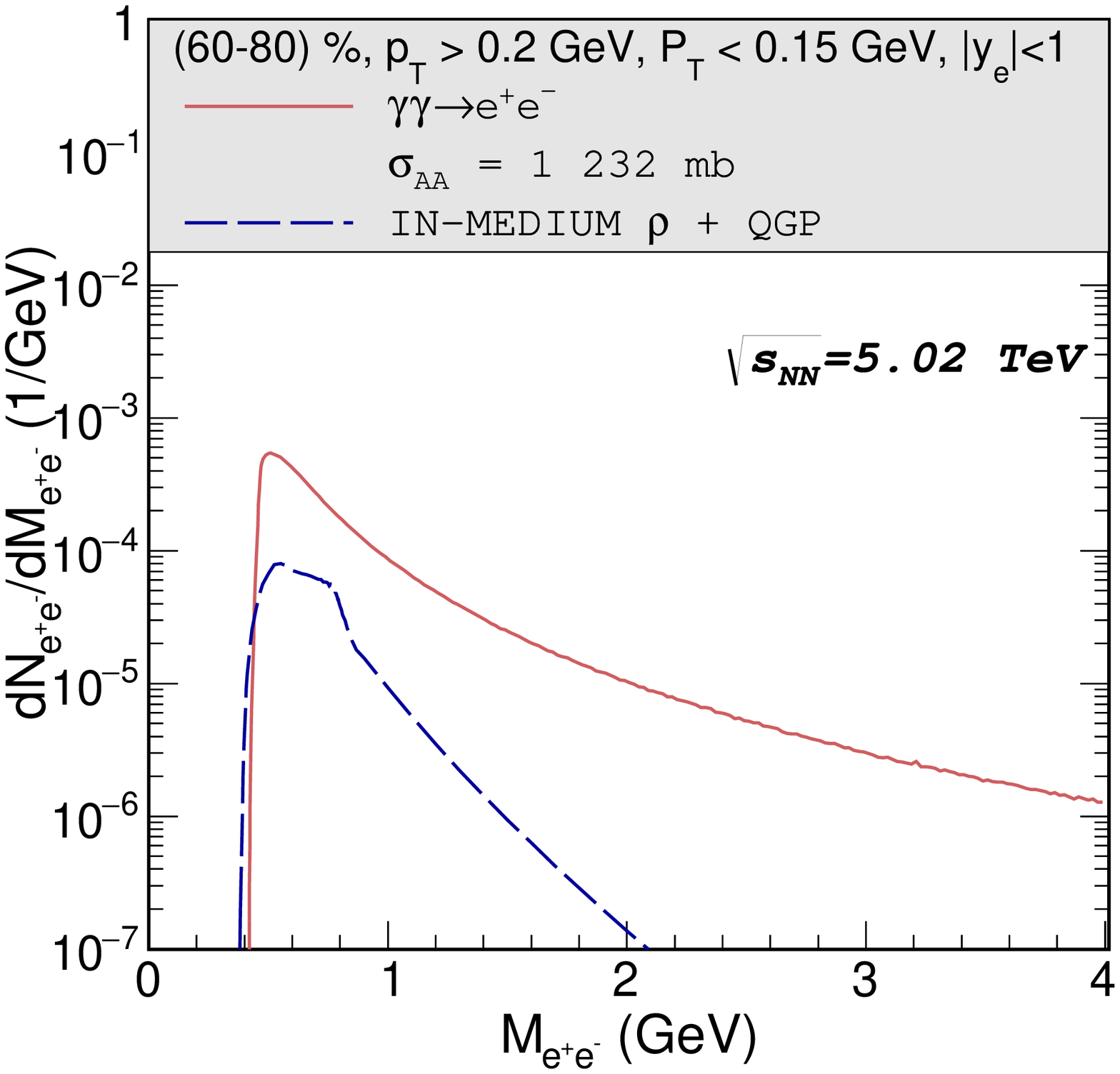}
	\caption{Our predictions for low-$P_T$ dilepton radiation in Pb+Pb ($\sqrt{s_{NN}}$=5.02\,TeV)
collisions from coherent $\gamma\gamma$ fusion (solid lines) and thermal radiation 
(dashed lines) for three centrality classes and acceptance cuts as specified in the 
figures.
	}
\label{fig:ALICE}
\end{figure}

The rather different relative importance of low-$P_T$ dilepton emission
from coherent $\gamma\gamma$ and thermal radiation sources found at RHIC and SPS
energies calls for their calculation at LHC energies. This is shown in 
Fig.~\ref{fig:ALICE} where we compare our predictions for the two sources for Pb+Pb 
collisions at $\sqrt{s_{NN}}$=5.02\,TeV for the same centrality classes and 
single-lepton acceptance cuts as for our RHIC calculations above. Compared to
the latter, the picture is qualitatively similar, although the strength of thermal
 contribution is relatively stronger, especially in semi-peripheral and central 
collisions where it is comparable and even larger, respectively,  than the 
$\gamma\gamma$ yield at low mass. 

\subsection{Excitation Function}
\label{ssec_excit}
\begin{figure}[!t]
	\includegraphics[scale=0.45]{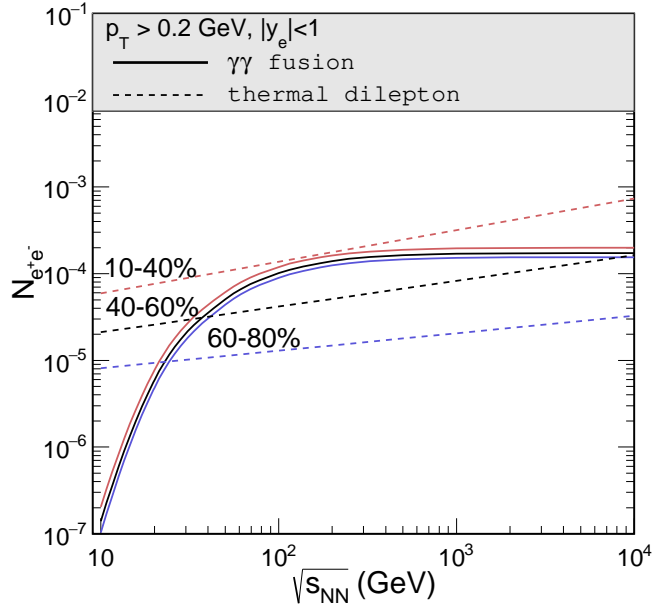}
	\caption{Excitation function of low-$P_T$ ($<$0.15\,GeV) dilepton yields from $\gamma \gamma$ fusion (solid lines) and 
thermal radiation (dashed lines) in collisions of heavy nuclei 
(A$\simeq$200) around midrapidity in three centrality classes, 
including single-$e^\pm$ acceptance cuts.
	}
\label{fig:sig_tot_sqrts}
\end{figure}

The results reported for the three different collision energies at the SPS, 
RHIC and the LHC in the previous three sections are now generalized into a systematic excitation 
function. In Fig.~\ref{fig:sig_tot_sqrts} we show the invariant-mass integrated  
low-$P_T$ (<0.15\,GeV) dilepton yields for the $\gamma \gamma$  and thermal components as a function of collision energy for 
the 3 centralities as used at RHIC and the LHC above, and 
including the same single-electron acceptance cuts. The 
photoproduction yields can be straightforwardly obtained from a 
direct calculation. For the thermal radiation, this would be much more involved. 
Instead we make use of the power-law like $N_{\rm ch}$ dependence of thermal 
radiation~\cite{Rapp:2013nxa} together with a power-law behavior of $dN_{\rm ch}/dy$ 
on collision energy~\cite{Basu:2016dmo} with the ansatz  
for $N_{ee}^{\rm th} = N_0 s^\beta$. We then fit the two parameters, $N_0$ and $\beta$, 
independently for each of the three centrality classes 
considered above, including the experimental acceptance cuts.  

The $\gamma \gamma$ fusion yield rises rather strongly in the tens of GeV collision 
energy regime, followed by a saturation above $\sqrt{s_{NN}}$ = 100 GeV, while thermal 
radiation shows a much more gradual increase, cf.~Fig.~\ref{fig:sig_tot_sqrts}. The latter 
therefore dominates above the former at low energies, $\sqrt{s_{NN}}\lesssim  20$\,GeV 
(as found in Sec.~\ref{ssec_sps}) at all centralities, and then increasingly so again in 
the TeV energy range for more central collisions. On the other hand, the $\gamma \gamma$ 
fusion contribution is most significant around the regime where it levels off, where
its yield is (much) larger than the one from thermal radiation 
for semi-/peripheral collisions, and  comparable for 
semi-central collisions. Our analysis therefore identifies the 
RHIC energy regime as the most promising ground to investigate 
this production mechanism.

\section{Conclusions}
\label{sec_concl}

We have studied low-$P_T$ dilepton production in ultrarelativistic heavy-ion collisions, 
by conducting systematic comparisons of the two sources that are believed to be prevalent 
in this regime, \ie, thermal radiation and photon-photon fusion within the coherent fields
of the incoming nuclei. The former was taken from a well-tested model including in-medium 
hadronic and QGP emission rates, while the latter was calculated utilizing photon fluxes 
with realistic nuclear form factors including the case of nuclear overlap. We first reconfirmed the finding of a recent 
STAR analysis that the combination of the 
two sources (augmented by a contribution from the hadronic final-state decay cocktail) 
gives a fair description of low-$P_T$ dilepton data in
Au-Au ($\sqrt{s_{NN}}$=200\,GeV) collisions in three centrality classes
for invariant masses from threshold to 4\,GeV (with the exception of the $J/\psi$ peak, 
indicating an additional production mechanism not included here). The coherent emission 
was found to be dominant for the two peripheral samples, and comparable to the cocktail 
and thermal radiation yields in semi-central collisions. 


At the lower SPS energies ($\sqrt{s_{NN}}$=17.3\,GeV) we found that the $\gamma \gamma$ 
contribution is subleading. Specifically, for acceptance-corrected low-$P_T$ dimuon 
spectra as measured by NA60 in MB In-In collisions, it reaches up to ten percent of 
the thermal radiation for masses near the dimuon threshold, rapidly falling off with 
increasing mass.  
On the other hand, at the high-energy frontier, the situation turned out to be 
similar to RHIC energies, although the role of thermal radiation relative to the
coherent mechanism is somewhat more pronounced in the multiple-TeV range. The 
interplay of these processes at the LHC is of particular interest in view of 
plans by ALICE~\cite{Antinori:2018} to lower the single-electron $p_T$ cuts and 
measure very-low mass spectra to possibly extract the EM conductivity from thermal 
emission. 

We have summarized our results in an excitation function of low-$P_T$ radiation covering 
three orders of magnitude in collision energy. While coherent production increases rather 
sharply, and then levels off, near $\sqrt{s_{NN}}$$\simeq$100\,GeV, thermal radiation increases 
more gradually with $s_{NN}$. This explains why the latter is dominant at the SPS, 
the former dominates at RHIC, and the latter becomes more important again at the LHC.

\acknowledgments
This work has been supported by the US National Science 
Foundation under grant no. PHY-1614484 (RR) and by the 
Polish National Science Center grant DEC-2014/15/B/ST2/02528 (MKG, WS, AS). 


\end{document}